# Design of a dynamic model of genes with multiple autonomous regulatory modules by evolution *in silico*

Alexander V. Spirov

Dept. of Applied Mathematics and Statistics, The State University of New York at Stony Brook, Stony Brook NY 11794-3600, USA
Email: spirov@kruppel.ams.sunysb.edu

New approach to design a dynamic model of genes with multiple autonomous regulatory modules by evolution in silico is proposed. The approach is based on Genetic Algorithms, enforced by new crossover operators, especially worked out for these purposes. The approach exploits the subbasin-portal architecture of the fitness functions suitable for this kind of evolutionary modeling. The effectiveness of the approach is demonstrated on a series of benchmark tests.

## 1. Introduction

Classical understanding of the mechanisms behind biological evolution served as the inspirational model for an entire order of heuristic optimization techniques, known in general as EC. In the past decade, research in molecular biology and genetics has conclusively shown that living organisms successfully utilize biomolecular implementations of EC for effective solving of problems in survival and adaptation. The most obvious examples of this type would be the mechanisms of antibody selection in a higher organism's adaptive immune system [1], and their counterparts, the mechanisms of antigen variability in pathogenic organisms, such as viruses and bacteria [2;3]. The advantage of such systems is that they are based on principles and mechanisms that seem similar to biological evolution [4;5;6]. Unlike biological evolution, however, they are also conducive to observation and experimentation. The principles of their function may be then fully defined in terms of EC and implemented on a computer [10-13].

Recent studies have brought to light the means by which the natural world carries out evolutionary search. *Natural* GA acts as a somewhat flexible hybrid optimization technique, used both in higher and lower organisms, albeit in differing ways. Specifically, our approach to EC is characterized by the use of operators that implement reproduction and diversification of genetic material in a manner inspired by the mechanisms of retroviral recombination [7;8] and the genetic-engineering technique known as DNA shuffling [9]. We will refer to our technique as *Retroviral* Genetic Algorithms or *retroGA*. Although related topics such as immune system modeling draw great attention from the field [10;11;12;13;14], there has to date been no work done on the algorithms of retroviral recombination or DNA shuffling.

RetroGA has many applications to problems of forced molecular evolution and has demonstrated impressive effectiveness on a series of benchmark tests. We selected these tests on the basis of their potential similarity to real-world problems of *in vitro* evolution and molecular-biological evolution. We gave special attention to the fitness functions as formal models, closely resembling real-world problems of molecular evolution. Some of the simplest fitness functions that demonstrate the properties of neutral subbasins linked by narrow pathways are the Royal Road (RR) and Royal Staircase (RS) fitness functions.

Recent publications of van Nimwegen with co-authors [15;16;17;18;19] emphasized the population dynamics of various RR and RS fitness functions. According to these authors, RR & RS problems often exhibit "evolutionary stasis", time periods when essentially no change takes place in population fitness. Stasis is one of the most interesting features of these functions because it is also frequently observed in both natural evolution and in evolutionary computation. In fact, van Nimwegen with co-authors draws attention to RR functions as a model of natural evolution.

It is becoming clear that the dynamics of evolutionary processes on fitness landscapes with neutrality are qualitatively very different from evolutionary dynamics on rugged landscapes [15-18]. A major impetus for this work is the lack of suitable models and theory for such landscapes. Common perception of landscape structure (multi-modal or rugged) in the GA literature can be inapplicable for optimization of the class of evolutionary scenarios that we deal with in this communication.

Our intent is to develop this approach to the point where it becomes possible to clearly gauge its utility to practical implications of its possible implementations, such as directed in vitro molecular evolution and biomolecular computation.

It is well known that natural evolution has successfully solved a great many very complicated bioengineering, biochemical, biomechanical, and bio-design problems. Many of these are still beyond our own power to solve. New findings in evolutionary biology, molecular biology, and functional genomics have, however, shed some light on the approaches used by Nature as inventor and designer.

*Biomolecular Computing*: The notable similarity between computational evolution in Artificial Life and EC, and molecular evolution *in vitro* has drawn the attention of researchers since the early 1990s [20]. The possibilities of this approach in comparison to those of "silicon-based technologies" have been widely discussed in the literature [21;22]. The current state-of-the-art in genetic engineering is now capable of developing this interesting topic to the level of actual *in vitro* experiment. Recently, a team of researchers set before themselves the goal of developing an *in vitro* DNA implementation of GA [23;24;25;26;27;28;29;30;31;32]. They have already defined several possible advantages of the *in vitro* GA implementation, and proposed a concrete design for a DNA based GA for some known benchmark problems: the MAX 1s Problem, the Royal Road (RR; See P. 9) Problem, and the "Cold War" Problem [24;25;29-32]. The team has made progress in developing procedures and protocols for an effective implementation of these problems in an experimental context. However, moving forward in this direction has brought to light significant problems, particularly in the methods used to physically separate DNA molecules in accordance with the RR fitness functions [23;27;28].

*Directed (Forced) Molecular Evolution:* In the past decade, experiments in *in vitro* molecular evolution ceased to be mere bench-top curiosities and have become a full-fledged track in current biotechnology [33;34;35]. The success of this track has been driven to a significant degree by the discovery of procedures that simulate recombination *in vitro* such as error-prone PCR [36;37] DNA shuffling [9;38], staggered extension process (StEP) recombination [39] restriction-ligation [40], the microgene method [41], and Y-ligation [42].

A typical round of *in vitro* evolution involves challenging the molecules to perform a specific task, then isolating and selectively amplifying the functional molecules. These procedures usually are carried out in a stepwise manner [34]. However studies of the pathways leading to functional sequences may be facilitated by the *continuous in vitro evolution* strategy [43;44;45;46]. In this system catalytic selection and amplification occur in the same reaction vessel, allowing many generations of selection to be conducted with maximal speed and minimal effort.

## 2. The Approach

### 2.1. Natural GA

The molecular machines that rearrange DNA often process molecules according to certain signal sequences. From a computational point of view, these are analogous to marks or tags on a string. Molecular machines read these tags and interpret them as instructions for further string operations. Of the genetic diversification mechanisms that utilize such signal sequences, one of the most simple and well-known is retroviral recombination.

*Retroviral Recombination:* Recombination is the process by which progeny receive an arrangement of genes that is different from that of either parent [1]. The life cycle of retroviruses is characterized by the alternate use of DNA and RNA as genetic material [1;7;8]. Each viral particle entering a host cell contains two or more copies of the viral genome in RNA form. The next stage of the infection cycle that holds interest for us is the synthesis of a single DNA molecule from these two or more molecules of viral RNA.

This task is carried out by an enzyme called retroviral *reverse transcriptase* [47], which is directed by a multitude of signal sequences in the original RNA. As the transcriptase synthesizes the replica from its template, it may pass over one of these signal sequences. When it does so, the transcriptase releases the current template strand and shifts to a different one. These jumps (or template switches or strand transfers) are the key event of retroviral recombination [7]. The signal sequences that trigger template switches may be either breaks in the RNA molecule or pause sites (regions of the RNA molecule with a certain sequence that slows down the synthesis of the replica) [7].

There are genetic evidences indicating that the human AIDS virus (HIV) originated from the chance reshuffling of genes from two monkey viruses in chimpanzees. Chimpanzees are the primary reservoir for HIV. Retrovirus HIV is a descendant of simian (monkey) immunodeficiency virus (SIV).

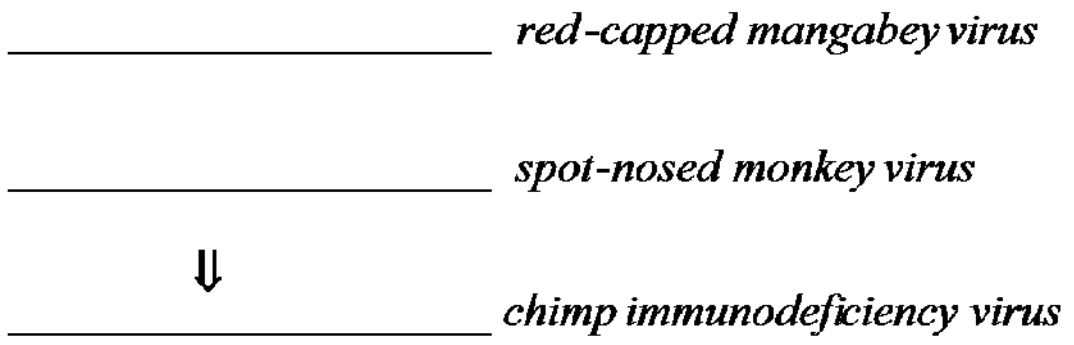

One part of the chimp virus resembled the virus found in the red-capped mangabey. Another resembled the virus from the greater spot-nosed monkey [48].

The hybridization took place inside chimps that had become infected with both strains of SIV after hunting and killing the two species of monkey.

It came to our attention that a generalization of this mechanism in genetic engineering could serve as the template for a powerful genetic diversification algorithm.

*Generalization of Retroviral Recombination:* These techniques are DNA shuffling (Sex PCR in particular), and Random-Priming Recombination (RPR) [9;49]. They are based on Polymerase Chain Reaction (PCR) and may be described as homology-based PCR. DNA shuffling involves the enzymatic cleavage of a collection of related genes to a pool of random DNA fragments [9;49]. These fragments can be reassembled into full-length chimerical genes by repeated cycles of self-priming PCR: the fragments prime each other based on local homology, and recombination occurs when fragments from one copy of a gene prime on another copy, causing a template switch.

In Sex PCR, as in retroviral recombination, there are two types of signals that need to be present in nucleic acids. The first are the sites, or signals, of replication interruption. It is believed that a retrovirus uses breaks in the molecule for this task, as well as pause sites [7]. However, Sex PCR only uses breaks for this purpose. In order for the process of generating the replica to continue, the reverse transcriptase in complex with the incomplete replica must find the target site on the other molecule. In the case of retroviral recombination, this site certainly exists, as two or more homologous molecules take part in the process of replication. In the case of Sex PCR, the acceptor molecule need not be entirely homologous, but it must have at least a small region homologous to a corresponding region on the donor molecule. Consequentially, Sex PCR brings to light the other class of signal sites: sites of local homology.

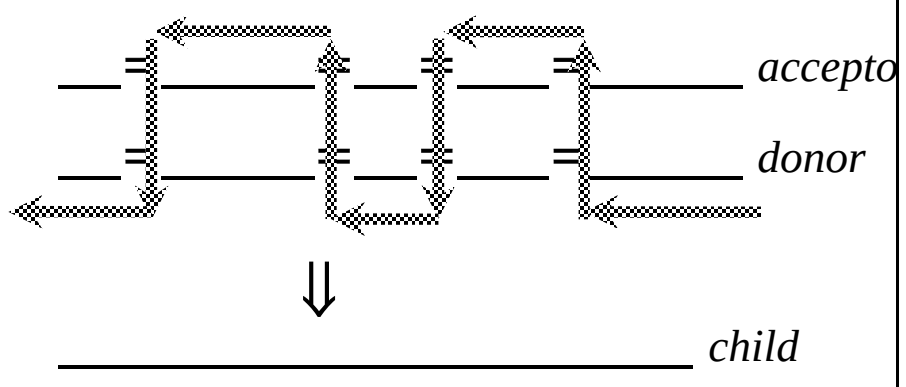

**Fig. 1.** Template switching during the RC operator action. The regions of local homology are denoted on both string as '_=_'.

## 2.2. The retroGA technique

Our technique is characterized by the use of operators that implement the reproduction and diversification of genetic material in a manner inspired by retroviral reproduction (RC operator) and a genetic-engineering technique known as DNA shuffling (GRC operator). These are our primary genetic operators as an alternative crossover and mutation operators. In everything else, our approach follows classical GA.

*The Reproduction/ Crossover operator:* The RC operator generates a child string from a given parent pair, combining the function of reproduction and crossover (Fig. 1). The pair of parents is selected, as in standard GA, by one of several predetermined strategies: *truncation*, *roulette-wheel*, etc. One string is selected as a donor, and another as an acceptor. Their sequences are then compared going from right to left for a short distance $l + \sigma$ (where $l < L$, L is the length of the whole sequence, $\sigma$ is a random integer, $\sigma \in 0,q$; $q<l$). If the required zone of local homology is not found, another couple is selected. If, and only if, a zone of complete homology (identity) of a size no less than S symbols ($S<q$) is, replica generation is initiated, and takes place in the first N symbols of the donor, from the first element to the last element of the found region of local homology (length of S symbols). The process then jumps onto the string of the acceptor. If next zone of local homology of a size no less then S is found, the replica generation is continued. The process then jumps onto the donor string.

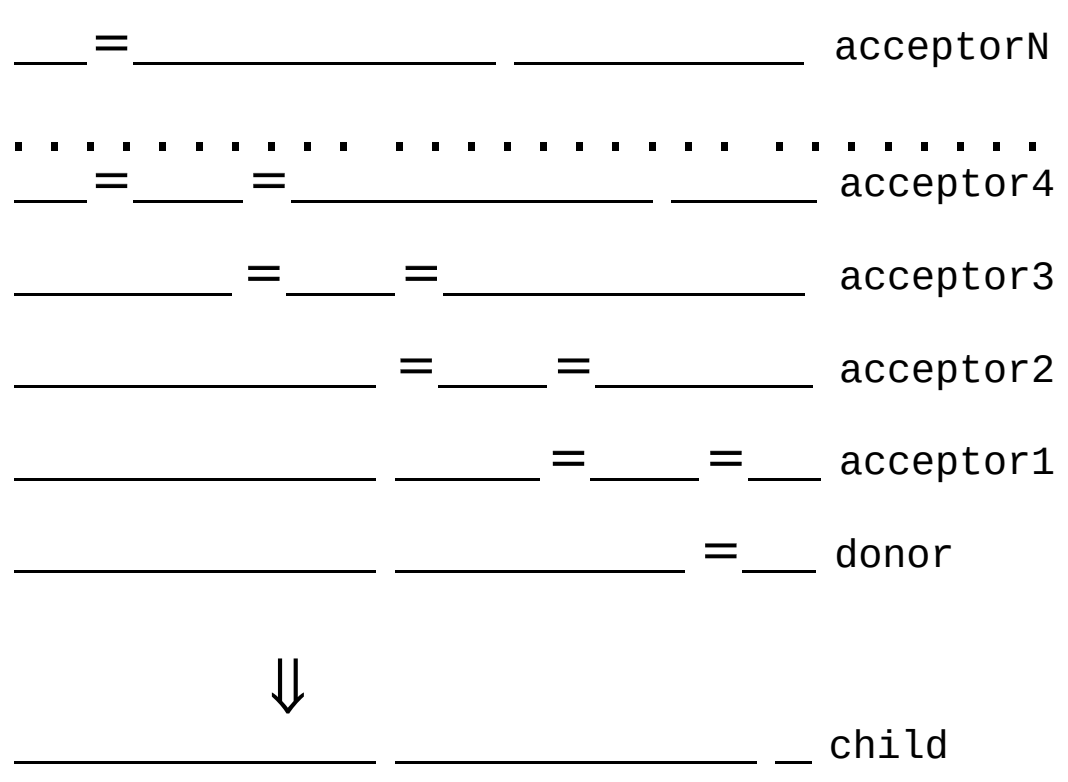

**Fig. 2.** Principle of the GRC operator. The process of creating the child sequence by the GRC operator starts with the parent-donor. The process then jumps onto the string of the second parent (acceptor1) with the condition that it have at least a small region that is homologous (identical) with the donor in the region of the jump (denoted on both string as ‘_=_’). After a while, the process again jumps, this time from acceptor 1 onto acceptor 2, again with the condition that there exists a region of complete local homology between them. It is thus possible to describe acceptor2 as the third parent of the child sequence. The process of jumping from acceptor to acceptor continues until the creation of the child sequence from the sequences of its many parents is complete.

Afterwards, another search for complete local homology takes place between acceptor and donor. This process is iteratively repeated until the replica (child) is completed or no more homology region is found. The number of iterations is at most *t*. If replica is not completed and no more homologous region is found, the next candidate pair of parents is selected.

As was discussed above, homology-based PCR techniques may be naturally interpreted as a generalization of retroviral recombination processes. This inspired us to develop a generalization of the RC operator - the Generalized Replication/Crossover operator (GRC operator), which simulates the basic properties of the Sex PCR technique. It works in the following manner: A first pair of parent candidates is selected according to a predetermined selection strategy – the donor and acceptor1 (Fig. 2). Their sequences are then compared going from right to left for a short distance $l + \sigma$ (where $l < L$, L is the length of the whole sequence, σ is a random integer, $\sigma \in 0,q$; $q<l$). If the required zone of local homology is not found, another candidate for acceptor1 is selected. The number of attempts to find a suitable acceptor is at most *t*. If, and only if, a zone of complete homology of a size no less than S symbols ($S<q$) is found during an attempt to scan two sequences, do these two sequences become the donor and acceptor1 pair. Replica generation is then initiated, and takes place in the first N symbols of the donor, from the first element to the last element of the region homologous between the two parents. Afterwards, the acceptor2 candidates are selected, and a search for local homology takes place between acceptor1 and the putative acceptor2. If no such region is found, the next candidate is searched. This process is iteratively repeated until the replica (child) is completed, or until the *t* limit is exceeded.

### 2.3. Fitness Functions to Study Hard Evolutionary Problems

There is every reason to believe that both biological evolution and natural GA solve problems of considerable difficulty [4-6]. In current literature dealing with biological and artificial evolution, one can find various grades and classifications of difficult problems having to do with biology [50]. It is these benchmark problems that we will focus on below.

In many combinatorial optimization problems as well as in biological molecular evolution, we can use the “building block” (BB) hypothesis [51;52]. This hypothesis states that a solution can be decomposed into a number of BBs, which can be searched for independently and afterwards be combined to obtain a good or even optimal solution. The remaining part of this paper is substantially based on the BB hypothesis.

At the same time as the idea of rugged landscapes was gaining momentum, an alternative concept was also being developed. This idea was based on the hypothesis of substantial degeneracy in the genotype-to-phenotype and the phenotype-to-fitness mappings. The history of this idea dates back to Kimura [53], who argued that on the genotypic level, most genetic variation occurring in evolution is adaptively neutral with respect to phenotype. During neutral evolution, different genotypes in a population fall into a relatively small number of distinct fitness classes, each consisting of a set of genotypes with approximately equal fitness. In other words, the genotype space decomposes into a set of subbasins of isofit genotypes that are entangled with each other in a complicated fashion. Through neutral mutations, individuals walk randomly in a given subbasin, until one of them discovers a connection to a subbasin of higher fitness. The internal structure of a subbasin may be described as a *neutral network*, wherein states of identical fitness are interconnected in a complex fashion.

*Benchmark Problems to Simulate Real Molecular Evolution:* The aforementioned class of fitness functions with subbasin-portal architecture has already found a practical application in analyzing the evolution of the secondary structure of RNA [67-70]. It has been shown *in silico* that a population of evolved molecules in a simulated flow reactor does not move over a neutral network in an entirely random fashion, but instead tends to concentrate at highly connected parts of the network, resulting in phenotypes (secondary structures) that are relatively robust against mutations [66]. The tendency to evolve toward highly connected parts of the network is independent of evolutionary parameters (mutation rate, selection advantage, and population size) and is solely determined by the network's topology. Hence, one can infer properties of the neutral network's topology from simple population statistics [66].

Another example of molecular evolution with a predominance of neutral drift that was widely discussed in the scientific literature is the evolution of the regulatory regions of genes. Several characteristics of regulatory regions, such as functional redundancy, modular architecture, and low sequence specificity of transcription factor (TF) for their binding sites (BS), suggest that they have freedom to change without drastic changes in their functions [54;55;56]. The commonly held conception is that most single mutations in regulatory regions represent a very small contribution to phenotypic variation, allowing the fixation of weak deleterious mutations by genetic drift [57;58]. Such an application of Kimura's ideas on the phenotypically close to neutral character of many (or the majority of) point mutations [65] to the evolution of regulatory regions/modules may be naturally modeled by fitness functions that exhibit subbasin-portal architecture. A more general case, where a weak deleterious (maladaptive) mutation in a single site of a regulatory region is compensated by a mutation in a different region (compensatory neutral mutations [59]) is also assumed in the evolution of regulatory regions. Dynamics of this kind have features of a random walk on a neutral network, but lead to quantitative changes in the form of the turnover of functional sites over time, while maintaining the function of the regulatory region.

In the next few years, analogous work on the modeling of directed evolution may be carried out with small molecules of DNA and oligopeptides. Both a sufficient experimental basis and the programming tools to support it exist to allow this [60;61;62].

### 2.3.1. Rugged Landscapes

Wright's [63] creation of the "adaptive landscape" metaphor has had a strong effect on the theoretical analysis of evolutionary processes. The point of view that a typical combinatorial optimization and biological evolution fitness function may be modeled by rugged landscapes has gained considerable currency in recent years [64]. This fitness function is such that typically, even two points in the search space that are very close together have considerable differences in fitness. Correspondingly, it is accepted that such landscapes have a high number of local extremes, as well as difficult elements such as plateaus and deep valleys. According to this outlook, evolving populations typically get stuck on one of the local peaks. The probability of going from the current local maximum to a neighboring local maximum is low, since peaks are typically separated by deep valleys. It is even less enthusing to consider the probability of having the population find the global maximum. This model clearly illustrates the problems of evolutionary search in the living world.

### 2.3.2. Subbasin-portal Architecture

At the same time as the idea of rugged landscapes was gaining momentum, an alternative concept was also being developed. This idea was based on the hypothesis of substantial degeneracy in the genotype-to-phenotype and the phenotype-to-fitness mappings. The history of this idea dates back to Kimura [65], who argued that on the genotypic level, most genetic variation occurring in evolution is adaptively neutral with respect to phenotype. During neutral evolution, different genotypes in a population fall into a relatively small number of distinct fitness classes, each consisting of a set of genotypes with approximately equal fitness. In other words, the genotype space decomposes into a set of subbasins of isofit genotypes that are entangled with each other in a complicated fashion (Fig. 3). This means that although the fitness landscape might be rugged, there are always neutral ridges along which the genotype can move without affecting fitness. In some cases local optima might disappear completely from the fitness landscape, as in the RR fitness functions [19]. Through neutral mutations, genotypes walk randomly in a given subbasin, until one of them discovers a connection to a subbasin of higher fitness. The internal structure of a subbasin may be described as a *neutral network*, wherein states of identical fitness are interconnected in a complex fashion.

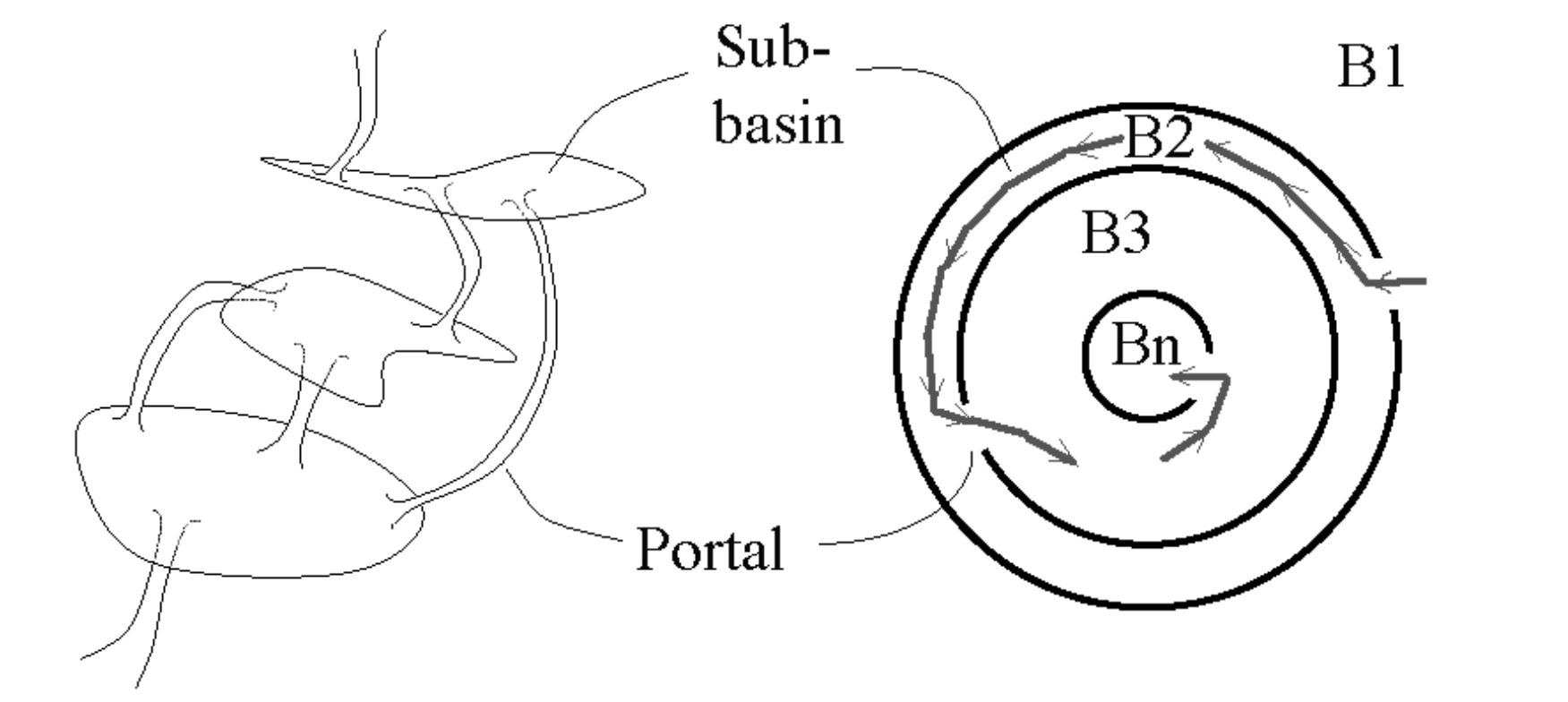

**Fig. 3.** General idea of a subbasin and portal architecture (left) and the dimensional hierarchy of subbasin and portals for the Royal Road / Royal Staircase fitness functions (right). (After [16]).

The aforementioned class of fitness functions with subbasin-portal architecture has already found a practical application in analyzing the evolution of the secondary structure of RNA [66;67;68;69;70].

*Royal Road Fitness Functions:* It was van Nimwegen with co-authors, who draw attention to RR as a model of natural evolution [15-19]. It is notable that some of the simplest fitness functions that demonstrate the properties of neutral subbasins linked by narrow pathways are the RR fitness functions. These functions were specifically proposed for testing the BB hypothesis, and whether recombination actually manipulated such BBs in the way that traditional GA theory assumed [71;72;73;74]. As a consequence of their formal simplicity, theoretical analysis can be carried out on the effectiveness of the simplest versions of GA and non-GA techniques for these functions. Despite the fact that they were intended as Royal Roads for GA, they in reality brought to light the substantial weaknesses of GA, caused first and foremost by the crossover operator. During the search for fitness functions that were simple for GA and difficult for other, non-evolutionary methods, a whole family of RR fitness functions were proposed, namely R1, R2, R3, and R4 [71]. Recently, elaborations such as the RS and Terraced Labyrinth Fitness Functions were introduced [15-18]. All of these functions demonstrate the neutral subbasin architecture. The difficulty of the RR functions increases from R1 to R4. Not one current optimization technique is capable of effectively dealing with the R4 fitness function.

The function **R1** is computed very simply: a bit string x gets 8 points added to its fitness for each of the given order-8 schemas of which it is an instance:

```
s1 = 11111111********************************************************; c1 = 8
s2 = ********11111111************************************************; c2 = 8
.....................................................................................
s7 = ************************************************11111111********; c7 = 8
s8 = ********************************************************11111111; c8 = 8
sopt= 1111111111111111111111111111111111111111111111111111111111111111; copt= 64.
```

The value R1(x) is the sum of the coefficients $c_s$ corresponding to each given schema of which x is an instance. Here, $c_s$ is equal to order(s). The fitness contribution from an intermediate stepping stone (such as the combination of $s_1$ and $s_8$) is thus a linear combination of the fitness contribution of the lower level components. This fitness function is an example of the class of functions with the subbasin and portal architecture (Fig. 3). The genotype space consists of all bit-strings of length 64 and contains 9 neutral subbasins of fitness 0, 8, 16, 24, 32, 40, 48, 56 and 64. There is only one sequence with fitness 64, 255 strings with fitness 56, 65534 strings with fitness 48, etc.

In the case of the second function **R2**, the fitness contributions of certain intermediate stepping stones are much higher. R2(x) is computed in the same way as R1, by summing the coefficients $c_s$ corresponding to each of the given schemas of which x is an instance.

The **R3** function differs from **R2** by the addition of spacers between BBs. In this case, the optimal string has the form of:

$s_{opt}$=11111111********11111111********11111111*******11111111********11111111********11111111********11111111********11111111,

where an * indicates a random bit and spacer sequences have no impact on the score. The R2 and R3 functions have the same subbasin-portal architecture as the R1.

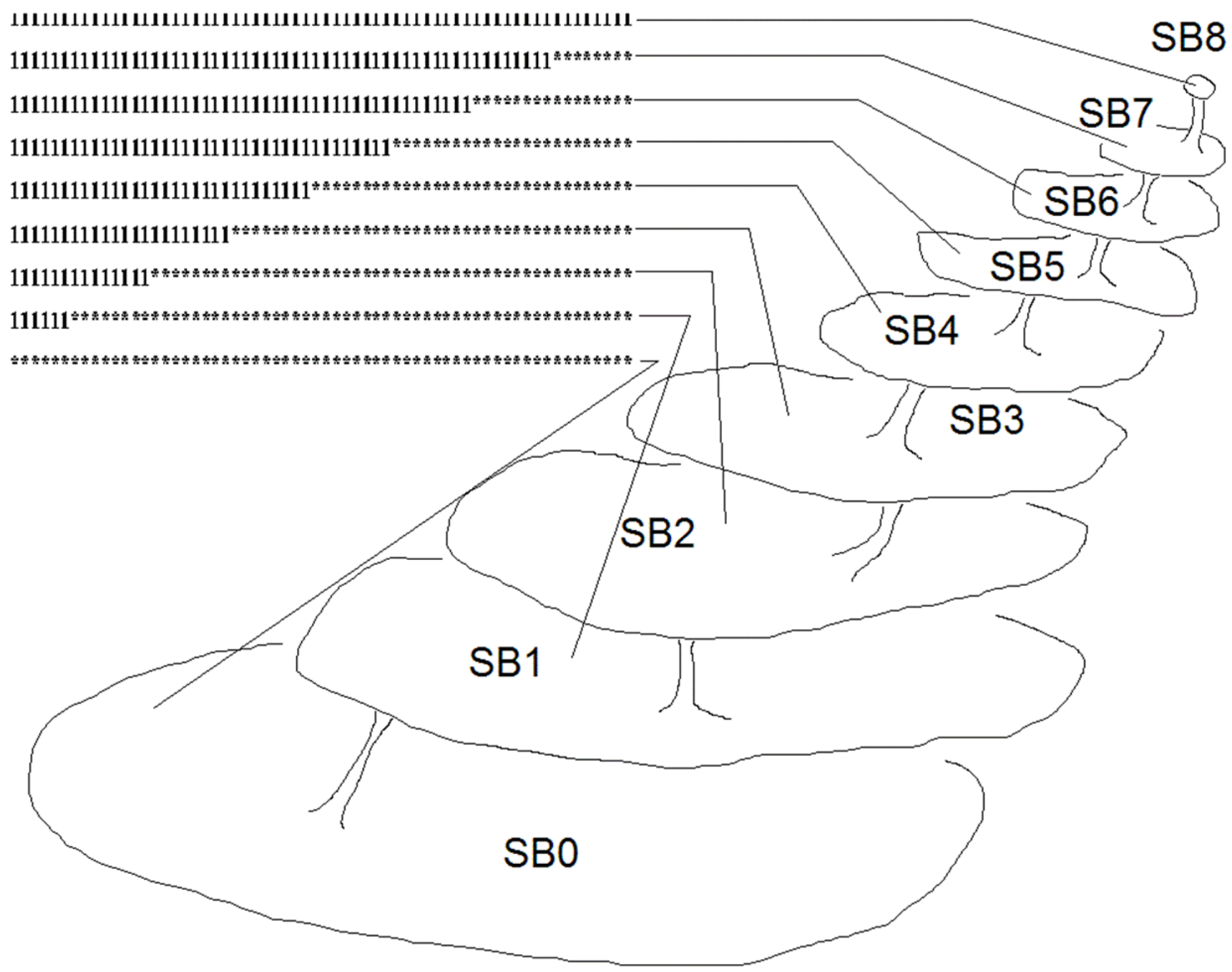

In the case of R1, there is only one string with fitness 64, 255 strings with fitness 56, 65534 strings with fitness 48, etc. The genotype space consists of all bit-strings of length 64 and contains 9 neutral subbasins of fitness 0, 8, 16, 24, 32, 40, 48, 56 and 64.

The highest and most difficult RR function is **R4**. The great difficulty of this problem lies in the fact that the presence of one, two, or even four non-neighboring elementary (in our case 8-bit) BBs in the string gives the same exact score (Level 1). The score will not increase until a BB of a higher order is found - such as a pair of elementary BBs (i.e., a 16-bit BB (Level 2), two 8-bit BBs neighboring each other). A level 3 BB consists of 4 neighboring 8-bit BBs (32 bits in total). Level 4 BBs are 64-bit, composed of eight 8-bit elementary BBs. Level 5 BBs cannot be achieved by any technique in practice, but would in theory consist of 128 bits. The genotype space consists of all bit-strings of length 128 and contains 5 neutral subbasins (Levels 0, 1, 2, 3 and 4).

*Royal Staircase Fitness Functions:* These are a generalization of the RR functions for which the subbasin-portal architecture is expressed in a more explicit form [15-18]. For any genotype there is a certain subset of bits that are fitness-constrained. Mutations in any of the constrained bits lower an individual's fitness. All the other bits are considered free bits, in the sense that they may be changed without affecting fitness. Of all possible configurations of free bits, there is a small subset of portal configurations that lead to increased fitness. A portal consists of a subset of free bits, called a constellation, which is set to a particular "correct" configuration. When a constellation is set to a partial configuration, the fitness is increased and the constellation's bits become constrained.

A RS fitness function corresponds to a Terraced Labyrinth whose tree is a simple linear chain (See Fig. 3). The RS function that we used in this paper was defined in a manner similar to RR functions (same length of string, and same size of BB as R1 and R2). Specifically:

```
s1 = 11111111********************************************************; c1 = 2
s2 = 1111111111111111************************************************; c2 = 3
.....................................................................................
s6 = 111111111111111111111111111111111111111111111111****************; c6 = 7
s7 = 11111111111111111111111111111111111111111111111111111111********; c7 = 8
s_opt= 1111111111111111111111111111111111111111111111111111111111111111; c_opt = 9
```

This version of the RS was used in the work of van Nimwegen and Crutchfield [18]. We used it so as to be able to compare our results to theirs. The genotype space contains 9 neutral subbasins of fitness 1, 2, 3, 4, 5, 6, 7, 8 and 9, and it reminiscent the R1-R3 functions architecture.

*Trap Functions*: One of the simplest discrete analogues of fitness functions with many maxima are concatenated trap functions [75;76]. They have been proven to be GA hard and are of particular interest from an experimental point of view for testing algorithm improvements. Here we use fully deceptive trap functions [77]. The following shows how to define a trap function of order *k*:

$$f(x) = \begin{cases} 1 & \text{if } u(x) = k, \\ 1 - \dfrac{1 + u(x)}{k} & \text{otherwise} \end{cases}.$$

where *u(x)* counts the number of 1-bits in string *x*. A higher-dimensional function is made by concatenating *n* trap functions together. The bit-string's fitness was computed as the sum of the fitnesses of the *n* traps. The concatenated trap function has 2*n* local optima. The global optimum is a string of all ones.

## 3. The Results

As was noted above, it is believed that such simple fitness functions as RR and RS reflect to a great degree the significant properties of biological evolutionary search. These functions are well-studied and are sufficiently simple to permit statistical analysis, and the comparison of their theoretical results with the results of experimental runs. Both RR and RS function tests used the same suite of programs. This package utilizes either the RC operator, or the GRC operator.

The following parameters were fixed in all test runs: The size of population (2,000) and the percentage of the population permitted to reproduce (15%). The initial population generated at random. The truncation strategy of reproduction was used when copies of chromosomes with scores exceeding the average value replaced all chromosomes having a score less then the average. For RC tests, the RC operator action probability was 0.1 per pair of parents. The operator's parameters are l=8, q=56, $\sigma \in$ 0,q, S=5, N $\in$ 8,q, and t=3 (See Section 2.2). For GRC tests, the GRC operator action probability was 0.15 per pair of parents and the operator's parameters are l=8, q=56, $\sigma \in$ 0,q, S=5, N $\in$ 8,q, and t=128.

**Table 1.** Performance of our technique versus standard GA. Values indicate number of function evaluations needed to reach optimum, averaged over 1000 runs.

| Func-tion | TECHNIQUES | | | | |
|---|---|---|---|---|---|
| | Std. GA [72-74] | RHMC [73,74] | MGE technique [78-80] | RC operator | GRC operator |
| R1 | 61,334±2,304 | 6,179±186 | 32,920±14,450 | 81,795±35,836 | 18,210±5,418 |
| R2 | 73,563±1,794 | 6,551±212 | 32,209±10,312 | 101,986±32,671 | 15,449±18,163 |
| R3 | 75,599±2,697 | No data | 34,436±6,239 | 136,566±77,241 | 16,720±29,977 |
| R4, $4^{th}$ l. | 86,078±17,242 | 95,027±17,948 | 155,465±45,003 | - | 20,949±17,480 |
| R4, $5^{th}$ l. | - | - | 265,359±55,165 | - | 151,653±218,335 |
| RS | ~500,000 | No data | 269,871±63,288 | 122,855±66,459 | 152,630±129,750 |

'-' means that neither run reached this level within the maximum of $10^6$ function evaluation.

*Royal Road Fitness Functions*: We had four functions, R1-R4, that differed in regards to the effectiveness of various evolutionary and non-evolutionary techniques (Table 1).

Experiments with the RC operator showed that this version of our approach does not exceed the efficiency of standard GA in the case of RR functions.

Surprisingly, it was the GRC operator that ended up being the most effective of all the strategies tested. On every benchmark test (Table 1), it outperformed all others by a significant margin. Notably, its performance in the case of R1 approached the non-evolutionary Random-Mutation Hill-Climbing (RHMC, [72-73]) algorithm. Mitchell et al. characterizes this algorithm as the simplest version of Simulated Annealing [71]. It was only three times less effective than RHMC (or even two times, depending on operator parameters), while RHMC outperformed standard GA by a factor of 10.

The GRC operator achieved the fourth level of the R4 test in 98% of the runs, and the fifth level in 67%. Through the use of the GRC operator, we were able to find an evolutionary technique that was sufficiently effective on these classical problems that it was capable of outperforming standard GA by a factor of 3 to 4 on all RR functions. We received similar results when using an earlier version of this technique called the MGE (Mobile Genetic Elements) approach [78;79;80].

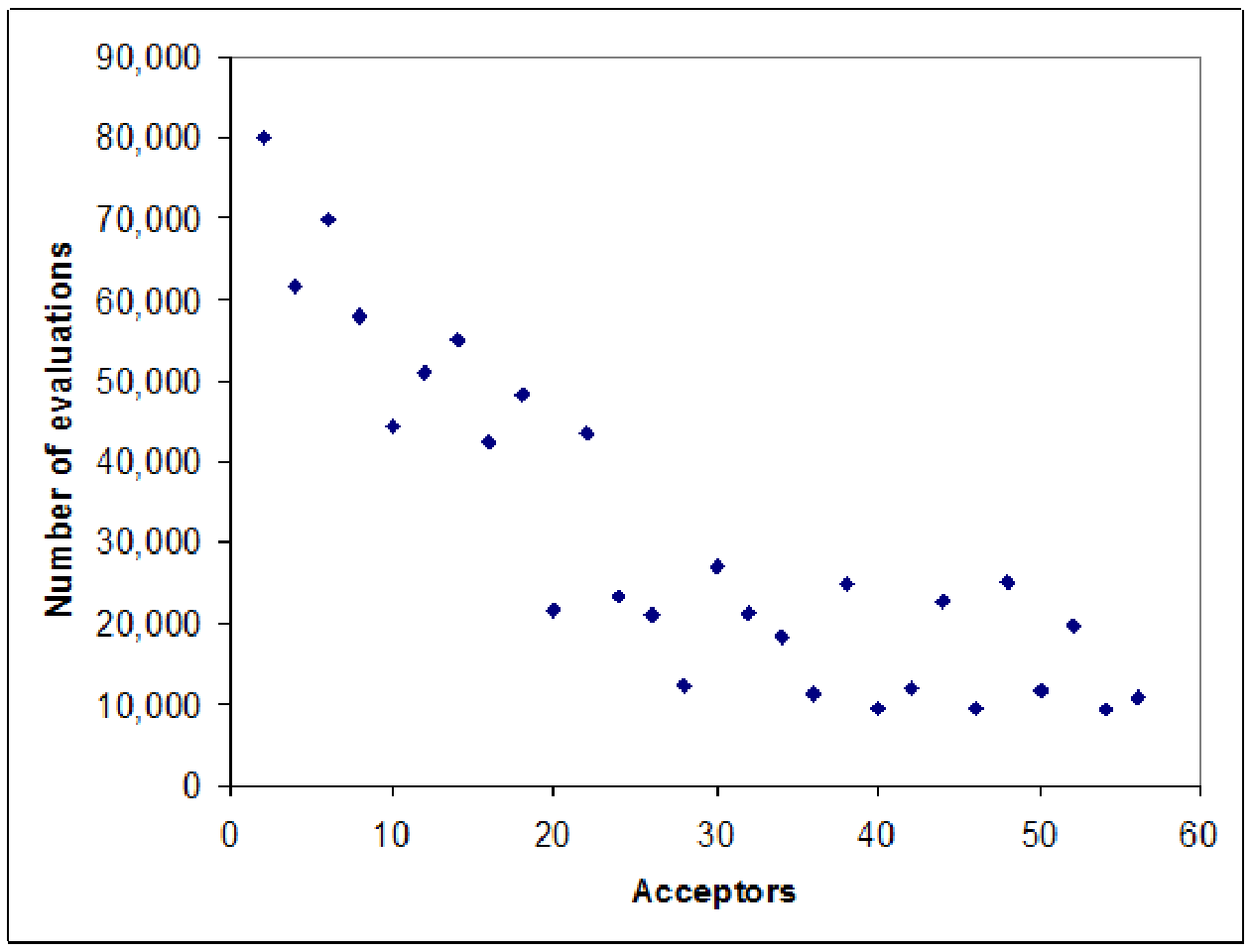

GRC outperforms RC: Performance of the GRC operator on the R1 fitness function against the limit of acceptors. Values indicate number of function evaluations needed to reach optimum, averaged over 100 runs. The efficacy of GRC operator depends on the number of acceptors.

*Royal Staircase Fitness Functions*: To our surprise it is the RC operator that found the answer to the RS test four times as fast on average than standard GA, while the GRC operator could do it more than three times as fast (Table 1).

*Approach to solve simulations of problems representative of in vitro evolution.* Improvement in the computational tools for bioinformatics and *in vitro* evolution has resulted in new capabilities for modeling molecular evolution. From all the possible problems in *in silico* evolution whose results we had sufficient confidence in, we selected the problem of simulating the forced evolution of transcriptional regulatory regions/modules. We would like to note, though, that our goal is not to produce simple conclusions on the evolutionary path of certain specific regulatory regions. Rather, our goal is to study the subbasin-portal architecture sequence space for some of well-studied regulatory modules and to compare the speed and effectiveness with which our methods solve these problems with that of methods that may be implemented in work on directed evolution. This would be an excellent chance to test the various methods that we developed and studied in earlier Aims of the project. More so, we will concentrate on those problems that we will mathematically analyze by the van Nimwegen and Crutchfield approach [15-19]. As such, we will have the chance to compare the results of numerical (computational) simulations with the results of analyzing the subbasin-portal architecture sequence space on these particular problems [97].

In order to carry out experiments in the forced evolution of regulatory regions *in silico*, we need quantitative methods of judging the distance of a given sequence to a known or desired one. Bioinformatics has in its purview a number of recognized methods for the *in silico* identification of transcriptional regulatory regions/modules [81]. We chose to utilize rule-based architectures as a method of recognizing clusters of TFBSs. Rule-based architectures build on studies of composite response elements [82;83]. This approach was already used successfully to detect and classify several different regulatory regions [83;84;85].

For our target sequence (solution), we have selected the sequence of the ribosomal RNA (rRNA) operon promoter rrn P1 in *E.coli*. The reason we chose this particular task is that prokaryotic promoters in general and these promoters in particular are without a doubt the most heavily studied [86]. In simulating directed evolution, the goal is to achieve the target sequence as rapidly as possible, starting from a pool of nucleic acid molecules of random sequence and length within predefined limits (150-200 bp). We use proven rule-based modeling algorithms and tools to produce a quantitative measurement of the distance between a given sequence and the regulatory region sequence desired [83-85]. These software packages are FastM / ModelCreator and ModelInspector, both free for non-commercial use [87]. Work begins with the creation of a model of a regulatory region based on some of its known properties. The model is created in FastM or ModelCreator. It requires a set of sequences containing a single type of regulatory unit (e.g. a set of homologous promoters from different species) and a very simple initial model (e.g. two TFBS and their sequential order). By means of iterative repeats the program refines and elaborates the initial model, adding new elements into it [85]. The final model of a regulatory region consists of descriptions of all individual elements, their mean scores, their sequential order, their relative frequencies, and the distance distributions observed in the training set.

The other tool, ModelInspector, uses this model to evaluate a given sequence. It does so by calculating the distance of the sequence from the determined model. This total element score and the score of element distances are then used to evaluate the fit of complex elements to the model. As an alternative method for developing a regulatory model to evaluate given sequences, we could, for instance, use the *Stubb* program, utilizing Hidden Markov Models [88].

*Simulation of the Directed (Forced) Evolution of Prokaryotic Promoters*: The core promoter of *E.coli* has a length of approximately 60 bp and is characterized by the presence of several conserved sites with spacers in-between. It is believed that while the sequence of these spacers is not significant, their length is of extreme importance [1]. There are at least four well-conserved features in a bacterial promoter: the starting point (usually ‘CAT’); the -10 sequence (‘TATAAT’ consensus); the -35 sequence (‘TTGACA’ consensus); and the distance between the -10 and -35 sequences. We will focus on the strongest type of *E.coli* promoters – the rRNA operon promoter rrn P1. Each rrn P1 promoter sequence contains an AT-rich sequence called the upstream (UP) element [89] upstream of the -35 element. UP elements increase transcription 20-to 50-fold [90]. Its consensus is AAA a/$_t$ a/$_t$ T a/$_t$ TTTT**AAAA, where * indicates a random base. In addition, three to five binding sites for the Fis protein (FisBS) increase transcription three- to eightfold [90;91]. The weight matrix for the binding site of this transcription factor has been defined [92]. Thus, the desired sequence is:

**[FisBS]**<~5 bp>**[FisBS]**<~5 bp>**[FisBS]**<~15 bp>**AAA a/$_t$ a/$_t$ T a/$_t$ TTTT**AAAA**<~4 bp>**TTGACA**<16-19 bp>**TATAAT**<5-9 bp>**CAT.**

Going by these facts, it is possible to interpret the evolution of the rrn P1 promoter as an example of evolution with the Royal Staircase fitness function. In other words, evolution could proceed by a route starting at the core promoter of reasonable sequence, to a far more powerful promoter with the UP element, to a promoter with maximal strength (another order of magnitude stronger) with a block of FisBSs. As such, this particular version of the Royal Staircase fitness function for the evolutionary search of the rrn P1 promoter has a familiar appearance (Cf. P. 11):

$s_1$ = ***********************************************…*************************************************TTGACA***…***TATAAT***…***CAT, $c_1 = \Delta$

$s_2$ = **************************…*****************************AAA a/$_t$ a/$_t$ T a/$_t$ TTTT**AAAA***…***TTGACA***…***TATAAT***…***CAT, $c_2 = \sim 35\ \Delta$

$s_3$ = *********************…*****************[FisBS]***…***AAA a/$_t$ a/$_t$ T a/$_t$ TTTT**AAAA***…***TTGACA***…***TATAAT***…***CAT, $c_3 = \sim 100\ \Delta$

……………………………………………………………………………………………………………………

$s_{opt}$ = ***[FisBS]***…***[FisBS]***…***[FisBS]***…***AAA a/$_t$ a/$_t$ T a/$_t$ TTTT**AAAA***…***TTGACA***…***TATAAT***…***CAT, $c_{opt} = \sim 150\ \Delta$

$\Delta$ is an arbitrary small coefficient (the values of coefficients $c_i$ will be established in the course of completing this Aim).

It would be very interesting to compare our expected results with the results obtained by simulating the forced evolution of an RNA molecule, the subbasin-portal architecture of whose sequence space has been already shown and statistically characterized [97].

There was, however, another reason for our decision to focus on the forced evolution of a prokaryotic promoter. The evolution of a simpler promoter (promoters from the T7, T3, and SP6 phages) has already been successfully implemented by continuous *in vitro* evolution [93]. As such, upon the completion of this project, we will have sufficient theoretical basis to plan future work in implementing a type of *in vitro* evolution that would permit us to test the effectiveness of our approach in a true bench top experiment.

## 4. Discussion and Conclusions

The mechanisms of diversification in natural GA are not analogous to mutation and crossover operators in computational GA. In computational GA, these mechanisms are global, act statistically upon the entire population, and use predetermined parameter values. In natural GA, however, the character of the mutation depends on the sequence of the given gene. It may be said that a gene contains not only information that is used to determine its fitness, but also instructions on how to mutate itself afterwards. As such, mutation operators in natural GA are local, and their action depends on the sequence of the particular gene in question.

Our conclusion was that there is a fundamental difference in the quality of the methods of artificial recombination implemented by the RC and GRC operators and standard GA's crossover operator. The positions of the sites of crossover and exchange between two strings in computational GA are chosen randomly. However, in biology, crossover occurs at sites of high homology between two molecules of nucleic acid. These regions of high homology may be naturally interpreted as BBs. As such, crossover operations in natural world do not destroy BBs, but instead conserve them wholly, while the material between the BBs undergoes crossover exchanges and point mutations (See Fig. 2). It is well-known that the destruction of already-discovered BBs by GA's crossover operator is one of the major problems of GA, and was originally brought to light by experiments with the RR fitness functions. Because of this, the capability of homology-based PCR techniques to conserve already located BBs is of tremendous interest to us.

It is hard to overestimate the significance of understanding and simulating biological evolutionary search mechanisms. We believe that methods of discrete optimization developed by the living world have significant meaning for interdisciplinary research. The new algorithms for evolutionary computation that we borrow from the living world are to a significant degree domain-independent. Because of this, they may be easily implemented in various EC techniques. Firstly, this has an impact on GA and GP: our approach's basic algorithms may be easily added to already-developed libraries. In particular, we refer here to the use of the RC and GRC operators.

In the past several decades, computational GA has become an effective mathematical instrument for modeling and analyzing the processes and mechanisms of biological evolution [9;94]. RetroGA has the potential to have the same effect on *in vitro* molecular evolution. Thus, further study and development of retroGA would serve to lay the foundation of a mathematical theory describing the processes and mechanisms behind the evolution of biological macromolecules *in vitro*.

It is perfectly reasonable to consider current techniques for selecting biological macromolecules with desired properties [95;96] as *in vitro* implementations of specialized variants of EC. This similarity consists of, firstly, the use of genetic engineering versions of the point mutation and crossover operators (Cf. [9;49]). Second, and more important to us, is the fact that these problems exhibit fitness functions that belong to the same class as the well-studied RR fitness functions. It is well-known that standard GA, utilizing only the point mutation and crossover operators, are insufficient for solving these types of problems. Quantitative mathematical analysis and numerical simulations have to this point only been carried out for a single problem in this type of directed evolution: the selection of RNA [97]. Despite noteworthy conclusions regarding the properties of that problem's fitness functions [67-70], no discussion has taken place regarding the adequacy of the methods used to diversify molecules before selection. In general, this field currently suffers from the lack of a theoretical basis for judging the effectiveness of various methods for diversifying nucleic acids. This becomes particularly evident in problems involving the selection of macromolecules with new properties that do not already exist in various precursors in natural world, i.e. from scratch [96;98;99].

## Acknowledgments

I thank Leonid Zamdborg for assistance with preparation of this manuscript and for useful comments on the MS. This work is supported by Joint NSF/NIGMS BioMath Program, 1-R01-GM072022-01.